\documentclass[conference]{IEEEtran}

\usepackage{cite}
\usepackage{amsmath,amssymb,amsfonts}
\usepackage{algorithmic}
\usepackage{graphicx}
\usepackage{textcomp}
\usepackage{xcolor}

\usepackage{tikz}
\usetikzlibrary{backgrounds}
\usepackage[outline]{contour}
\contourlength{2pt}

\definecolor{inp_clr}{rgb}{0.0,0.0,0.70}
\definecolor{out_clr}{rgb}{0.70,0.0,0.0}
\definecolor{res_clr}{rgb}{0.70,0.0,0.70}
\definecolor{inpe_clr}{rgb}{0.4,0.0,0.80}
\definecolor{oute_clr}{rgb}{0.70,0.0,0.5}
\definecolor{out_edge_clr}{rgb}{0.3,0.5,0.3}
\definecolor{inp_edge_clr}{rgb}{0.5,0.5,0.8}
\definecolor{res_edge_clr}{rgb}{0.5,0.5,0.5}

\def\BibTeX{{\rm B\kern-.05em{\sc i\kern-.025em b}\kern-.08em
    T\kern-.1667em\lower.7ex\hbox{E}\kern-.125emX}}

\newcommand\copyrighttext{
    \small \begin{center} \color{gray} \textcopyright\,2025 IEEE.  Personal use of this material is permitted.  Permission from IEEE must be obtained for all other uses, in any current or future media, including reprinting/republishing this material for advertising or promotional purposes, creating new collective works, for resale or redistribution to servers or lists, or reuse of any copyrighted component of this work in other works. \end{center}}
\newcommand\copyrightnotice{%obtained from https://tex.stackexchange.com/a/55815
    \begin{tikzpicture}[remember picture,overlay]
    \node[anchor=south,yshift=10pt] at (current page.south) {\parbox{\dimexpr\textwidth-\fboxsep-\fboxrule\relax}{\copyrighttext}};
    \end{tikzpicture}
    }
\begin{document}

\title{Dynamic Reservoir Computing \\ with Physical Neuromorphic Networks
}

\author{\IEEEauthorblockN{Anonymous Authors}}

\author{
    \IEEEauthorblockN{Yinhao Xu}
    \IEEEauthorblockA{\textit{School of Physics}\\
    \textit{University of Sydney}\\
    NSW, Australia\\
    yinhao.xu@sydney.edu.au}
\and
    \IEEEauthorblockN{Georg A. Gottwald}
    \IEEEauthorblockA{\textit{School of Mathematics and Statistics}\\
    \textit{University of Sydney}\\
    NSW, Australia\\
    georg.gottwald@sydney.edu.au}
\and
    \IEEEauthorblockN{Zdenka Kuncic}
    \IEEEauthorblockA{\textit{School of Physics}\\
    \textit{University of Sydney}\\
    NSW, Australia\\
    zdenka.kuncic@sydney.edu.au}
}

\maketitle
\copyrightnotice
\begin{abstract}
Reservoir Computing (RC) with physical systems requires an understanding of the underlying structure and internal dynamics of the specific physical reservoir.
In this study, physical nano--electronic networks with neuromorphic dynamics are investigated for their use as physical reservoirs in an RC framework.
These neuromorphic networks operate as dynamic reservoirs, with node activities in general coupled to the edge dynamics through nonlinear nano--electronic circuit elements, and the reservoir outputs influenced by the underlying network connectivity structure.
This study finds that networks with varying degrees of sparsity generate more useful nonlinear temporal outputs for dynamic RC compared to dense networks.
Dynamic RC is also tested on an autonomous multivariate chaotic time series prediction task with networks of varying densities, which revealed the importance of network sparsity in maintaining network activity and overall dynamics, that in turn enabled the learning of the chaotic Lorenz63 system's attractor behavior.
\end{abstract}

\section{Introduction}

Physical Reservoir Computing (RC) is an emerging computing paradigm that builds on and extends standard RC by exploiting physical systems as the reservoir \cite{Tanaka2019,Jaeger_2023,Stepney2024}. 
A particularly promising approach to physical RC is that of physical reservoirs with neuromorphic dynamics, such as neuron-like and/or synapse-like properties. 
Given the brain's efficient information processing abilities, as well as growing evidence that at least some brain regions operate in a similar way to RC \cite{Enel2016,Suzuki2023}, neuromorphic physical RC presents a compelling approach to realizing efficient machine learning.

While optical, mechanical and quantum reservoirs with neuromorphic dynamics continue to be explored, the most promising approach to date is with electronic reservoirs \cite{Liang2024}. 
Self-organized nano-electronic networks, comprised of nanoparticles or nanowires, are particularly promising because, in addition to exhibiting neuro-synaptic dynamics, they also naturally form a recurrent network structure with neuromorphic topology \cite{Loeffler2020}. 
Indeed, they exhibit a neuromorphic structure--function relation that is difficult to replicate with mathematical models of recurrent neural networks and neuronal dynamics \cite{Suarez2021}.

Several studies have demonstrated physical RC using neuromorphic networks comprised of different nano-electronic material substrates \cite{Vahl2024}. This study focuses on neuromorphic nanowire networks, typically comprised of metal-dielectric material such that cross-points undergo resistive memory (memristive) switching under an electrical bias due to electro--ionic transport across nanoscale junctions \cite{Stieg2012,Milano2019,Diaz-Alvarez2019,kuncicNeuromorphic2021}. 
In addition to synapse--like memristive dynamics, nanowire networks also exhibit neuron--like spiking dynamics, with statistical correlations that obey the criteria for avalanche criticality observed also in cortical neurons \cite{hochstetterAvalanches2021,Dunham2021}.

\begin{figure}[h]
\centering
\scalebox{1.15}{\input{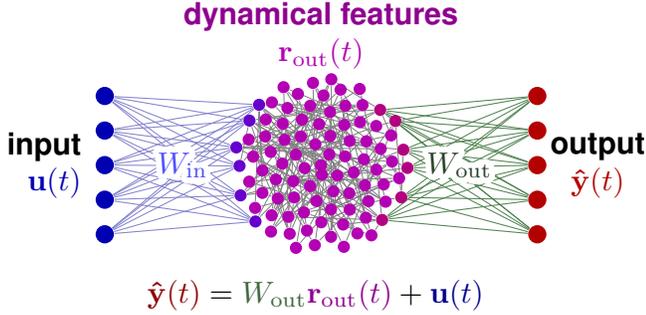}}
\caption{ 
    Schematic illustration of dynamic RC with a physical neuromorphic network. Input voltage signals $\mathbf{u}(t)$ weighted by $W_{\rm in}$ are delivered to a subset of nodes, while dynamical features $\mathbf{r_{\rm out}}(t)$ are read out from other nodes and used to train weights $W_{\rm out}$ in an output layer to generate $\mathbf{\hat{y}}(t)$ that best approximates the desired outputs $\mathbf{y}(t)$. 
    The physical neuromorphic network is typically a nano-electronic circuit with coupled nonlinear neuro-synaptic dynamics on both the nodes and edges, so in general, the reservoir weights are conductance-based and evolve in time, in contrast to standard RC. 
}
\label{fig:rc_schematic}
\end{figure}

Both experimental and simulation studies have shown that the nonlinear neuro-synaptic dynamics of nanowire networks can be exploited for physical RC for a range of learning tasks, including MNIST handwritten digit classification \cite{Milano2022,Zhu2023}, voice classification \cite{Lilak2021,Kotooka2024}, pattern recognition \cite{Milano2023_prc}, waveform regression \cite{Sillin2013,Fu2020,zhuInformation2021,Loeffler2021,Daniels2022}, as well as chaotic and highly nonlinear time series prediction \cite{kuncicNeuromorphic2020a,Zhu2020,Zhu2023_L2L}. 
At the same time, these studies reveal important differences between RC implemented with physical reservoirs and that of mathematical models of reservoirs. 
As nano-electronic systems, nanowire networks are constrained by the laws of physics (i.e Kirchhoff's conservation laws, memristor equation of state), which impose coupled dynamics on the nanowire nodes and memristive edges. 
This means that the reservoir dynamics are more complex than in standard RC, where nonlinear dynamics is introduced into the reservoir solely via nodes using a nonlinear mathematical activation function (e.g. tanh), with random fixed edge weights. 
This is in general not possible with a physical reservoir. 
Thus, physical RC in this context may be more aptly described as \emph{dynamic} RC: coupled nonlinear dynamics in both the nodes and edges generates dynamical features that can be used for learning (cf. Fig.~\ref{fig:rc_schematic}). 

The purpose of this study is to demonstrate dynamic RC via simulations of physical neuromorphic networks.
Section~\ref{sec:method} outlines the physically--motivated model used for the simulations and the RC set--up for a specific task, autonomous prediction of the Lorenz63 chaotic system. 
This is a common task in standard RC and its variations \cite{kosterDatainformed2023,pathakUsing2017,platt2022systematic,akiyamaComputational2022,mandal2025learning}, but has not previously been implemented using a physical neuromorphic reservoir. 
Section \ref{sec:struct} investigates the effects of the underlying physical network structure on its dynamical responses. 
In section \ref{sec:dynam}, a proof of concept of the neuromorphic network's dynamic RC capabilities is numerically demonstrated with the Lorenz63 multivariate chaotic time series prediction task.
As this study implements the simplest dynamic RC setup using a simulated network that is relatively small compared to physical neuromorphic networks (which can typically reach millions of physical nodes \cite{Sillin2013,Diaz-Alvarez2019}), no attempt is made to optimize parameters for achieving state-of-the-art RC results.
Rather, the results validate whether RC can be fully realized with internal dynamical responses of a physical neuromorphic reservoir, in place of using activation functions in an idealized mathematical reservoir.

\begin{figure*}[hbt]
	\centering
	\includegraphics[width=\linewidth]{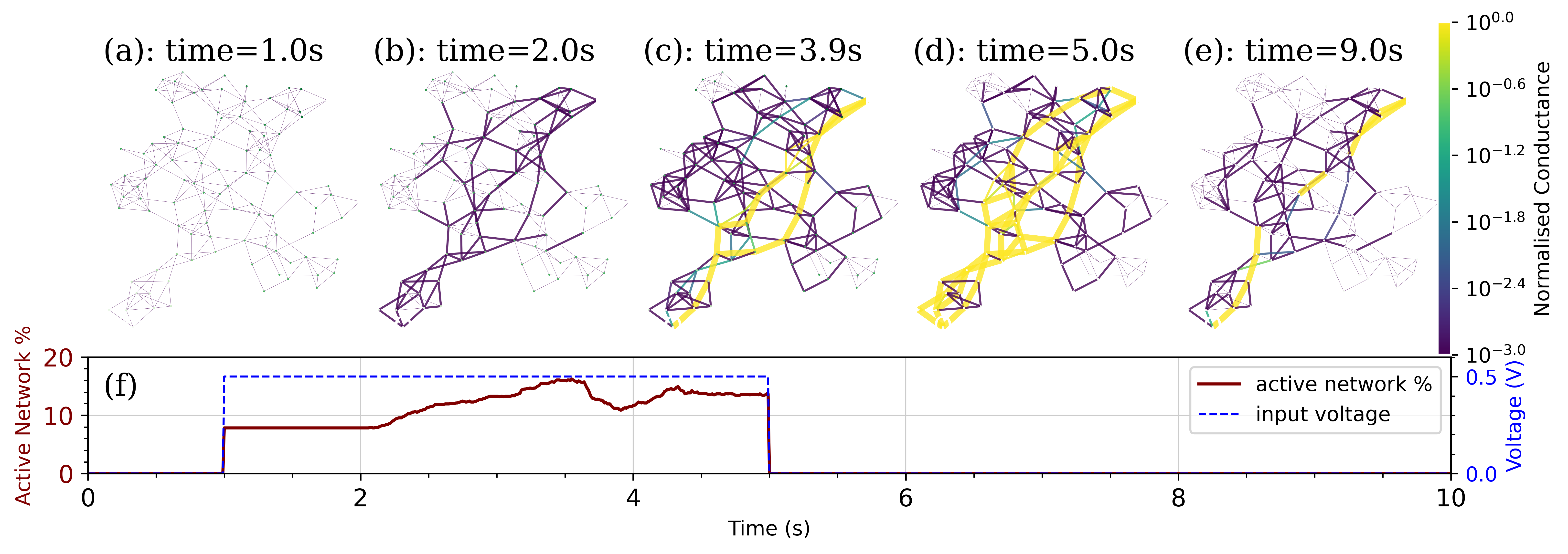}
	\caption{Graph visualizations of a simulated $100$-node, $261$-edge neuromorphic network showing dynamic connectivity in response to a 0.5\,V DC voltage pulse of 4\,s duration.
    Snapshots are shown at times (a) $1$\,s, (b) $2$\,s, (c) $3.9$\,s, (d) $5$\,s, and (e) $9$\,s, with the normalized conductance on edges indicated by the colorbar and the thickness of the respective edges. (f) Plots showing the input voltage pulse (blue) and the percentage of the network that is active (red) as a function of time.
    }
	\label{fig:snap1}
\end{figure*}

\section{Methods}\label{sec:method}

\subsection{Neuromorphic Network Model}\label{sec:nwn}

Physical neuromorphic networks were simulated based on a model for self-assembled nanowire networks that display complex neuro--synaptic dynamics under electrical stimulation, where electrical flow between nanowires is governed by nonlinear memristive junction dynamics \cite{Milano2019,kuncicNeuromorphic2021}.  
A neuromorphic network can be abstracted as a graph, where nodes represent nanowires and edges represent memristors. 
The physical model of neuromorphic nanowire networks has been described in previous studies \cite{kuncicNeuromorphic2020a,zhuInformation2021,hochstetterAvalanches2021,Baccetti2024};
briefly, the graph's connectivity and Kirchhoff's circuit laws produce a system of linear equations (with intrinsic nonlinear dependencies) that can be solved and updated at each time point.

The dynamical behavior of these networks is illustrated in Fig.~\ref{fig:snap1}, which depicts snapshots of conductance changes over time as a voltage pulse is applied to a single node. 
As the voltage distributes across the nodes based on their connectivity, edge conductance evolve and form localized branches (Fig.~\ref{fig:snap1}(b)). 
These branches further redistribute the voltages of the nodes, culminating in the formation of the first high conductance path (Fig.~\ref{fig:snap1}(c), highlighted in yellow) between the input and a ground node in the electrical circuit. 
With continued application of the input signal, additional parallel conductance paths emerge (Fig.~\ref{fig:snap1}(d)).
After the input pulse, the conductance paths gradually diminish over time (Fig.~\ref{fig:snap1}(e)) demonstrating the presence of fading memory. 

The network exhibits different dynamic activity in specific regions at varying times, based on its connectivity and input signal.
As shown in Fig.~\ref{fig:snap1}(d), network activity, defined by the proportion of edges where the voltage difference is above a threshold set value (0.01\,V used here) such that it allows for edge conductance evolution, gradually increases and peaks at around $20$\% and then subsequently declines by around $5$\%. 
After $3.6$\,s, the rate of conductance changes decrease in parallel with diminishing voltage differences between nodes.
With multiple input nodes and more diverse bipolar input signals, the network's behavior follows similar dynamic patterns, but with richer complexity in its response. 

\subsection{Dynamic Reservoir Computing Framework}
Given a time varying input signal $\mathbf{u}(t) \in \mathbb{R}^{N_{\rm u}}$, and a known target signal $\mathbf{y}(t)\in \mathbb{R}^{N_{\rm y}}$,
the neuromorphic network can be used as a dynamic reservoir to learn the dynamics of the target signal to produce $\mathbf{\hat{y}}(t)\in \mathbb{R}^{N_{\rm y}}$ which best approximates $\mathbf{y}(t)$ (cf. Fig.~\ref{fig:rc_schematic}).
The reservoir input value $\mathbf{r}_{\rm in}(t)\in \mathbb{R}^{N_{\rm in}}$ is coupled to the input signal $\mathbf{u}(t)$ via an input weight matrix $W_{\rm in} \in \mathbb{R}^{N_{\rm in} \times N_{\rm u}}$ and an input bias vector $\mathbf{b}_{\rm in} \in \mathbb{R}^{N_{\rm in}}$, with entries of both randomly sampled from a uniform distribution: 
\begin{equation}
	\mathbf{r}_{\rm in}(t)= W_{\rm in} \; \mathbf{u}(t) + \mathbf{b}_{\rm in} \; .
\end{equation}
The output estimate $\mathbf{\hat{y}}(t)$ is coupled to the reservoir readouts  $\mathbf{r}_{\rm out}(t)$ via an external output layer, which consists of an output weight matrix $W_{\rm out} \in \mathbb{R}^{N_{\rm y} \times N_{\rm out}}$ that is trained to learn the dynamics of $\mathbf{u}(t)$:
\begin{equation}
   \mathbf{\hat{y}}(t) = W_{\rm out}\; \mathbf{r}_{\rm out}(t)+ \mathbf{u}(t) \; .
\end{equation}
The $\mathbf{u}(t)$ component included here fulfills a similar role to skip connections in residual networks \cite{he2016deep}.
$W_{\rm out}$ is trained via ridge regression, with the Tikhonov regularization parameter value set to $10^{-6}$.

The reservoir here is a simulation of the physical neuromorphic network, described above. 
In this context, it is a complex high-dimensional dynamical system which nonlinearly transforms $N_{\rm in}$ dimensional input signals $\mathbf{r}_{\rm in}(t)$ into $N_{\rm out}$ dimensional readout signals $\mathbf{r}_{\rm out}(t)$. Here, $N_{\rm out} \geq N_{\rm in}$, with only a subset $N_{\rm in}$ of all available nodes used as input nodes and the remainder used as readout nodes.
In the context of machine learning, the readouts $\mathbf{r}_{\rm out}(t)$ act as dynamical feature embeddings that can be used to learn the nonlinear dynamics inherent in the input data. 

\subsection{Autonomous Chaotic Time Series Prediction}
Simulated neuromorphic networks are used within the dynamic RC framework (Fig.~\ref{fig:rc_schematic}) to perform autonomous prediction of the multivariate Lorenz63 nonlinear system \cite{lorenzDeterministic1963}:
\begin{align}
    \dot{y}_1 &= \sigma (y_2-y_1) \, , \nonumber \\
    \dot{y}_2 &= y_1(\rho - y_3) - y_2 \, , \label{eqn:Lorenz63} \\
    \dot{y}_3 &= y_1 y_2 - \beta y_3 \, . \nonumber
\end{align}
To produce chaotic dynamics, the parameters were set to $\sigma = 10$, $\rho = 28$ and $\beta = 8/3$.

The desired output signal $\mathbf{y}(t)$ is now the Lorenz63 system, discretized with a step size of $\Delta t=0.005\,$s.
The current estimate $\mathbf{\hat{y}}(t)$ is used to predict $\mathbf{\hat{y}}(t+\Delta t)$ at one time step ahead.
During training, the input signal is $\mathbf{u}(t) = \mathbf{y}(t-\Delta t)$, and during autonomous prediction, the prediction at previous step $\mathbf{\hat{y}}(t-\Delta t)$ is instead assigned as the input (i.e. $\mathbf{u}(t) = \mathbf{\hat{y}}(t-\Delta t)$) to enable auto-generation without external input.

The three discretized variables of the Lorenz63 system in (\ref{eqn:Lorenz63}) serve as the input signals after being normalized to a mean of $0$ and standard deviation of $1$, then multiplied by an appropriate voltage of $0.2 \, \rm{V}$ before feeding it into the simulated neuromorphic nano-electronic network.
A training duration of $27,000$ time steps was used.

To quantify the accuracy of short-term prediction, 
the relative forecast error was defined as 
\begin{equation}
    \mathcal{E}_f(t) = \frac{||\mathbf{y}(t) - \mathbf{\hat{y}}(t)||^2}{\langle||\mathbf{y} - \langle \mathbf{y} \rangle||^2\rangle},
    \label{eqn:err}
\end{equation}
where $\langle \cdot \rangle$ denotes the time average.
The forecast time $t_f$ was defined as the longest time such that the relative forecast error $\mathcal{E}_f(t_f)$ is less than a threshold $\theta$:
\begin{equation}
    t_f = \max{(\lambda_{\rm{max}}\,t|\mathcal{E}_f(t)\leq\theta)}.
    \label{eqn:tf}
\end{equation}
A threshold of $\theta=0.4$ was chosen to be comparable with results of similar setups \cite{kosterDatainformed2023,akiyamaComputational2022}. 
The forecast time $t_f$ is measured in units of Lyapunov time of the true Lorenz system (i.e. normalized by $\lambda_{\rm max}^{-1}$), where the largest Lyapunov exponent is $\lambda_{\rm max}=0.91$. 

\section{Results \& Discussion}
\subsection{Network Structure} \label{sec:struct}

\begin{figure}[ht]
	\centering
	\includegraphics[width=\linewidth]{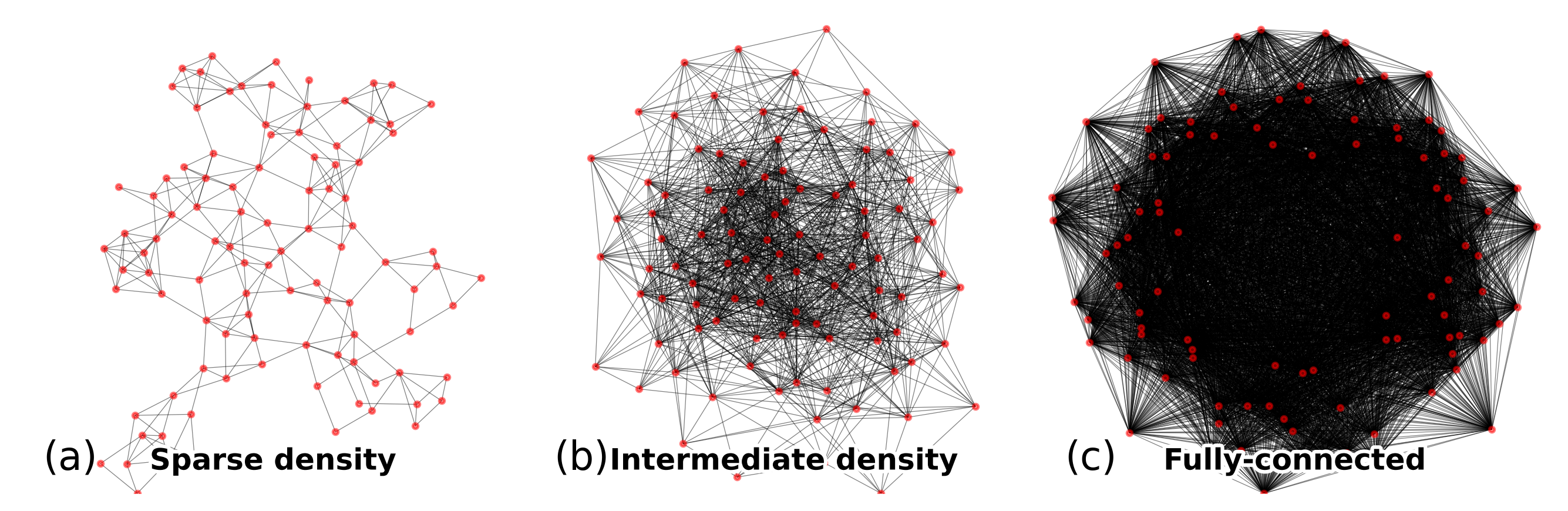}
	\caption{Graph representation of physical neuromorphic networks of varying densities: (a) a sparse, low density network with 100 nodes and 261 edges; (b) an intermediate density network with 100 nodes and 1517 edges; and (c) a fully--connected, high density network with 100 nodes and 4950 edges.}
	\label{fig:networks}
\end{figure}

Fig.~\ref{fig:networks} shows simulated physical neuromorphic nanowire networks of various network densities abstracted into their graph representations, with nodes (circles) representing nanowires and edges (links) representing their memristive junctions.
To investigate the influence of the underlying network structure on their response to time-varying input signals, the network size was chosen to be relatively small, with 100 nodes in all cases. 
The density varies in accordance with the number of edges and hence, connectivity. 

\begin{figure}[ht]
	\centering
	\includegraphics[width=\linewidth]{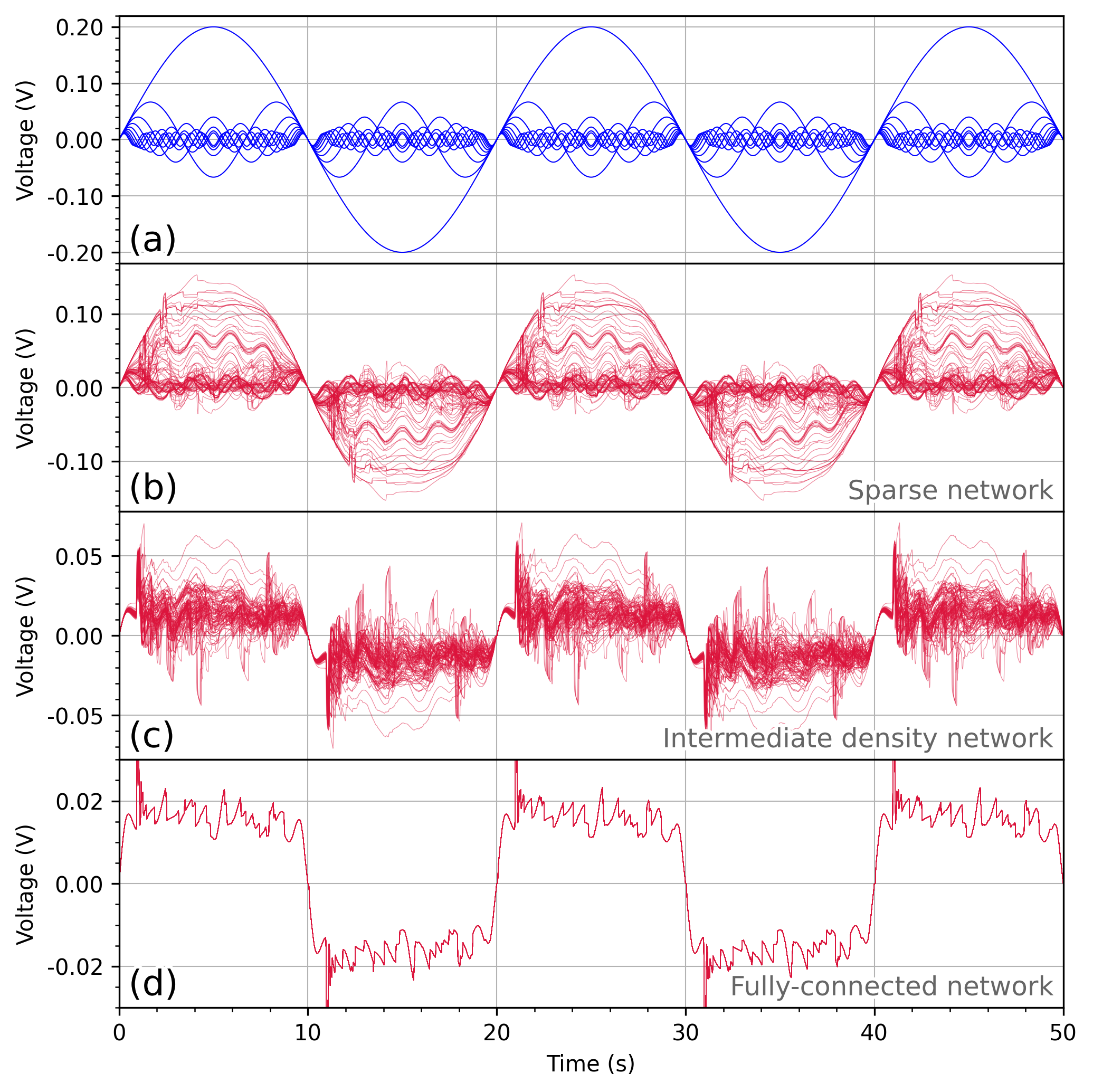}
	\caption{Neuromorphic network input--output mapping. (a) Input signals (blue, 10 Fourier modes of a square wave); and readouts (red) for (b) sparse, (c) intermediate density and (d) fully--connected networks. All simulated networks have 100 nodes, with 261, 1517 and 4950 edges for the sparse, intermediate density and fully-connected networks, respectively, as depicted graphically in Fig.~\ref{fig:networks}.
    }
	\label{fig:sine_readouts}
\end{figure}

Fig.~\ref{fig:sine_readouts} shows the input--output mapping of the simulated neuromorphic network reservoirs in Fig.~\ref{fig:networks}. 
The input signals, shown in Fig.~\ref{fig:sine_readouts}(a), are the first 10 terms of a periodic square wave's Fourier series, and each network reservoir was given the same 10 signals across 10 randomly selected input nodes. 
Fig.~\ref{fig:sine_readouts}(b) shows that output signals for the sparse network (Fig.~\ref{fig:networks}(a)) tend to be correlated with the input signals, with some appearing to be a linear combination of some inputs, while others effectively repeat the input signals.
Fig.~\ref{fig:sine_readouts}(c) shows that the readouts of a denser network no longer reflect the inputs and appear to exhibit more diverse nonlinear features compared to Fig.~\ref{fig:sine_readouts}(b). 
In stark contrast to this, the readouts of a fully connected network, shown in Fig.~\ref{fig:sine_readouts}(d), all overlap and are almost exactly the same; evidently all readouts are now the scaled sum of all inputs, as they generate a noisy square wave (the noise arises from the limited number of Fourier harmonics used as input, as well as the presence of nonlinearities from the memristive edge dynamics). 

These results demonstrate the importance of the underlying connectivity structure of physical neuromorphic network reservoirs; they are the foundation of a physical reservoir's dynamical response to inputs.
In conventional RC, the reservoir's dynamical response is determined by a nonlinear mathematical activation function applied to all the nodes, which have a fixed connectivity structure.
In physical nano-electronic reservoirs, such as the memristive networks considered here, node dynamics is instead coupled to and driven by the edge dynamics, which obey a memristor equation of state. 
The network connectivity structure imposes a physical limit on the collective network dynamics, which are constrained by Kirchhoff's electrical circuit laws. 
This means that the network connectivity influences how voltage inputs are continuously redistributed across the nodes and nodes do not need to be nearest neighbors to influence each other.
Due to their dynamics, physical neuromorphic networks can serve as effective reservoirs without being fully--connected, in contrast to conventional recurrent neural networks used as reservoirs.

The results in Fig.~\ref{fig:sine_readouts} can be explained in terms of network density, driven by the network dynamics of a complex electrical circuit.
In a sparse circuit such as Fig.~\ref{fig:networks}(a), voltage is distributed across relatively local nodes (i.e. nearest neighbors).
As such, the readout signal at any node location is approximately the input signals of the closest input nodes; this can be seen in Fig.~\ref{fig:sine_readouts}(b). 
In a circuit network of intermediate density, the voltage signal at every input node is distributed across more distant nodes, and thus influences more memristor circuit elements in a nontrivial way due to network heterogeneity.
In a highly dense circuit network (such as the fully--connected example in Fig.~\ref{fig:networks}(c)), the input voltage signal spreads indiscriminately across nodes such that memristor circuit elements have less potential difference available to activate and sustain their dynamics. 
This is why the readout signal at any node location is approximately a sum of every input signal, as is evident from the square wave in every readout signal in Fig.~\ref{fig:sine_readouts}(d). 

\subsection{Network Dynamics} \label{sec:dynam}

\begin{figure*}[ht]
	\centering
	\includegraphics[width=\linewidth]{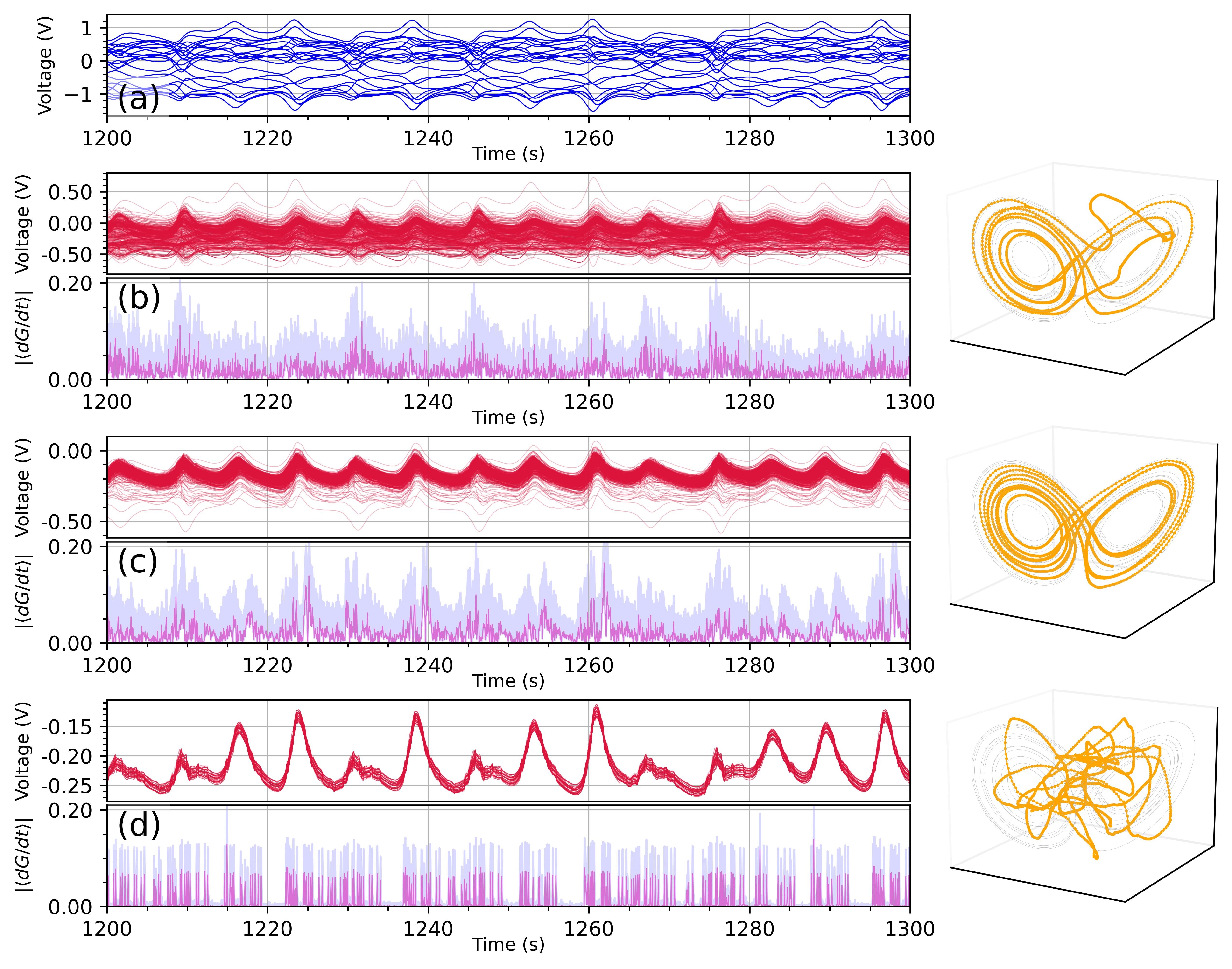}
	\caption{
    (a) 24 input signals (blue) comprised of a random linear combination of the three chaotic Lorenz system signals. (b)--(d) Node readout signals (red), amplitude of the change of conductance $dG/dt$ averaged over all memristive edges (magenta), its standard deviation (light blue shaded region), and the resulting autonomously forecasted Lorenz attractor (orange) compared to the actual attractor (gray), which commences at the first time step after training; 
    simulated networks are with 500-nodes of increasing density: 2119 edges, 9905 edges, and 123,671 edges, for (b)--(d) respectively.
    Time duration selected during the last stages of training.
    }
	\label{fig:lorenz_readouts}
\end{figure*}

The neuromorphic network's functionality as a dynamical reservoir was evaluated by assessing its performance on autonomous forecasting of the chaotic Lorenz system.
The input signals (Fig.~\ref{fig:lorenz_readouts}(a)) are a linear combination of the Lorenz signals \eqref{eqn:Lorenz63}. 
The dynamic reservoir readout signals (red) are shown in Fig.~\ref{fig:lorenz_readouts}(b)--(d) for 500-node networks of increasing density.
After training weights in an output layer with the readouts, the Lorenz system was autonomously generated without any external input to produce the attractors (orange) corresponding to each network density in Fig.~\ref{fig:lorenz_readouts}. 
Each plot also shows the corresponding average rate of change in edge conductance, $G$ (normalized), in the dynamical reservoirs. 

The readouts exhibit behaviors similar to that seen in Fig.~\ref{fig:sine_readouts} for smaller networks and simpler input signals. For the low density network (Fig.~\ref{fig:lorenz_readouts}(b)), the readouts exhibit some correlation with the input signals, while for the high density network (Fig.~\ref{fig:lorenz_readouts}(d)) the readouts overlap and are correlated with each other.
For the medium density network (Fig.~\ref{fig:lorenz_readouts}(c), the readouts also show correlations, but are interspersed with additional diverse features.
The results clearly show that this medium density dynamic reservoir is able to predict the Lorenz attractor better than the other two reservoirs.
Note that as a network becomes increasingly sparser, the average distance between readouts and input nodes increases, and hence, each readout becomes increasingly influenced by fewer input signals.
Effectively, each readout reflects the signal of the closest few input nodes, with some nonlinearities, reducing the overall effect of the network structure; in other words, for a sufficiently sparse network, similar readouts can be obtained without using a network, and can be approximately generated as just the scaled input nodes themselves.  

The rate of conductance changes in each dynamic reservoir also exhibits qualitative differences. 
The densest network (Fig.~\ref{fig:lorenz_readouts}(d)) exhibits large switching events,
with mean conductance frequently changing from low to high states,
whereas in the sparser networks (Fig.~\ref{fig:lorenz_readouts}(b) and (c)) the mean conductance changes between medium to high states.
To interpret these results, note that the densest network has 58 times as many memristive edges as the sparsest network;
since input signals are delivered with the same voltage amplitude across the same number of input nodes (ten), the significantly larger number of edges allows the signal to distribute much more uniformly across the network into all existing regions.
Thus, voltage differences between nodes in the densest network are considerably lower than in the sparser networks, the readout voltages are more similar, and the memristive edges are unable to evolve rapidly enough into meta-stable states like those achieved in the sparser networks.

\begin{figure}[ht]
	\centering
	\includegraphics[width=\linewidth]{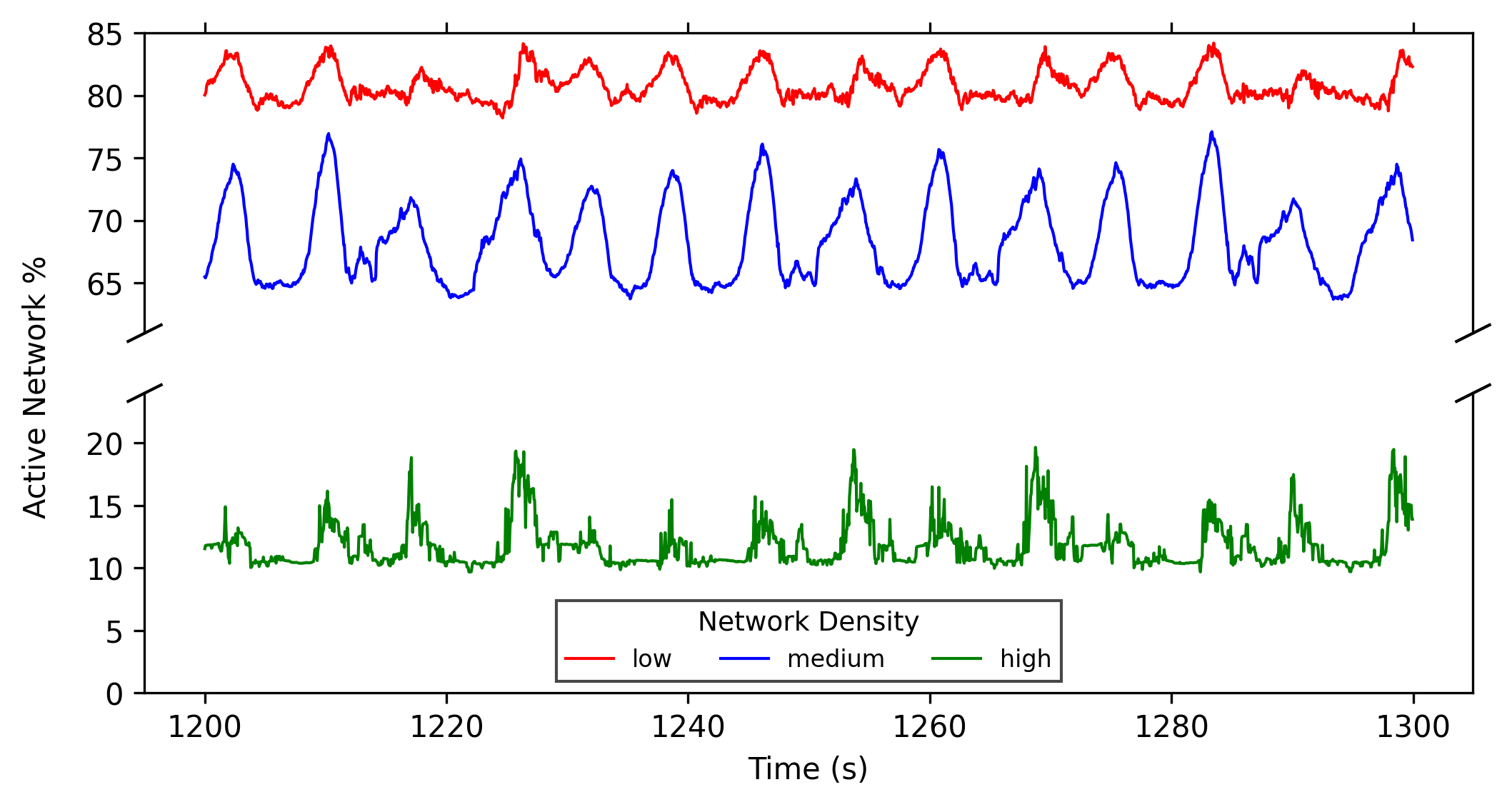}
	\caption{Percentage of network activation over time, for low (red), medium (blue) and high (green) density networks. 
    The network, task and time duration all correspond to that of Fig.~\ref{fig:lorenz_readouts}.}
	\label{fig:activecomps}
\end{figure}

This difference is highlighted further in Fig.~\ref{fig:activecomps}, which plots the percentage of network activation, defined by the ratio of memristive edges that have sufficient voltage bias to start evolving dynamically.
Only $15$\% of the densest network is active at any given time, whereas the sparser networks have higher and more comparable active components, with the lowest density network exhibiting a consistently high level of edge activity, at around $80$\%. 
Interestingly, the medium density network exhibits diverse high activity (ranging between 65--77\%), suggesting that \emph{dynamic activity} may be an essential key property needed for a dynamic reservoir to achieve high performance, at least in multivariate chaotic time series prediction tasks.

\begin{figure}[ht]
	\centering
	\includegraphics[width=\linewidth]{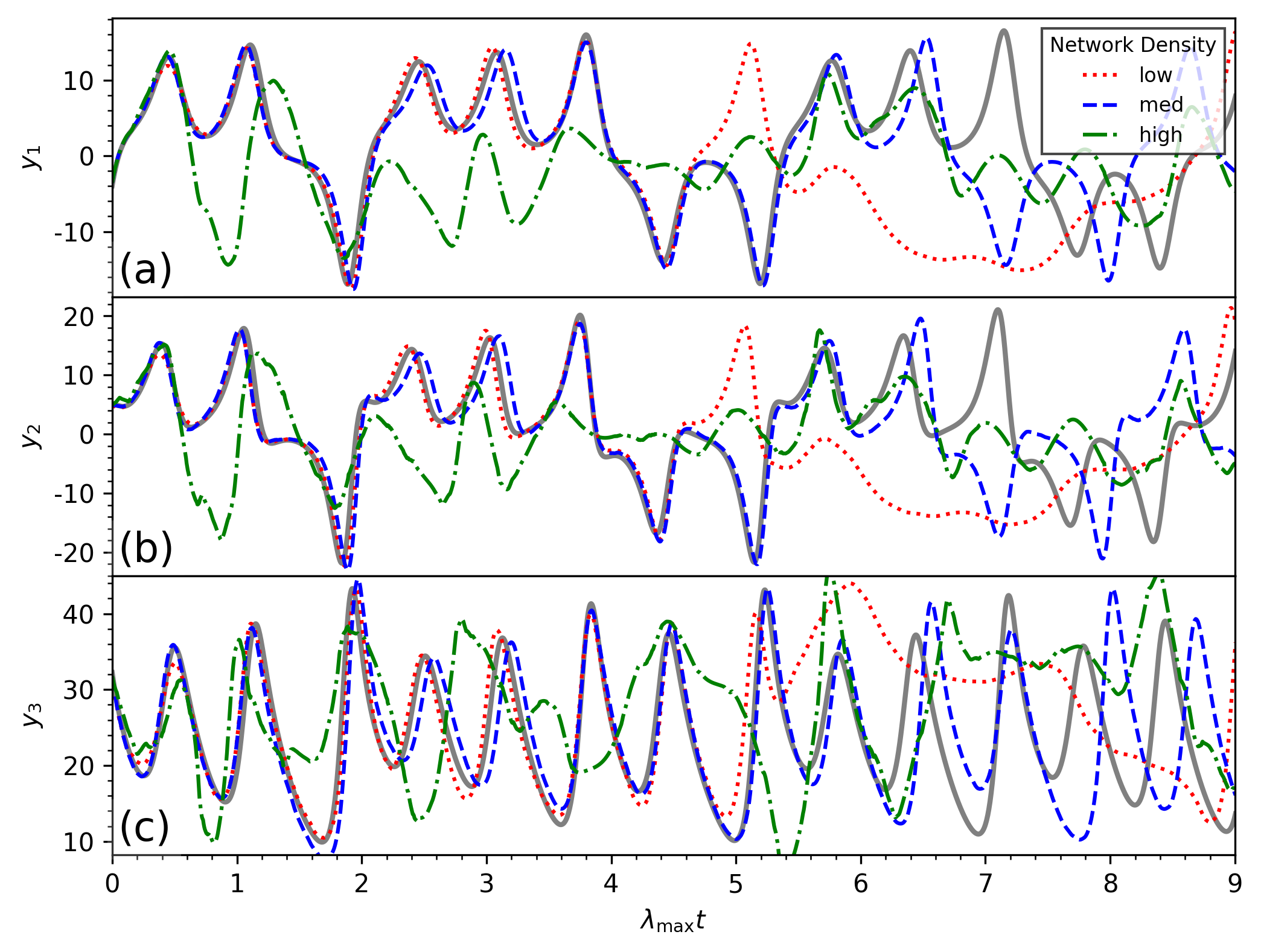}
	\caption{Autonomously predicted values of the Lorenz system. (a), (b) and (c) correspond to the $y_1(t)$, $y_2(t)$ and $y_3(t)$ variables of the Lorenz system \eqref{eqn:Lorenz63} respectively, with the prediction results from the low density 500-node 2119-edge network (red), medium density 9905-edge (blue) and high density 123,671 edge (green) networks, compared with the true Lorenz signal (black). 
    The predicted results here correspond to the attractors of Fig.~\ref{fig:lorenz_readouts}.
    The duration for forecasting is measured in units of Lyapunov times.} 
	\label{fig:pred}
\end{figure}
\begin{figure}[ht]
	\centering
	\includegraphics[width=\linewidth]{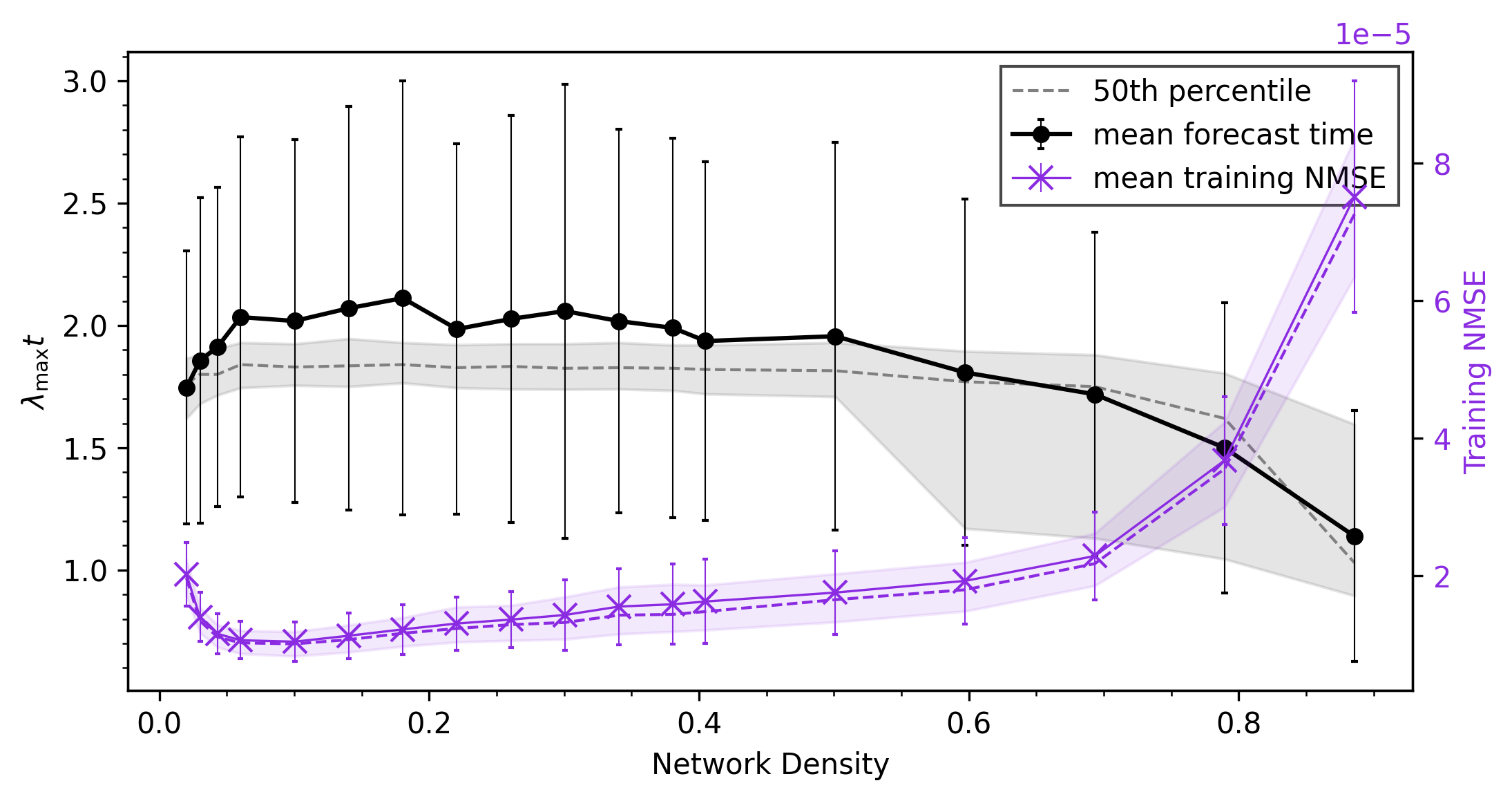}
	\caption{Average forecast time (black, in units of Lyapunov times) and the average normalized mean square error (purple) during training, with respect to network density, on 500 node networks. 
    The forecast error bars indicate one standard deviation, and the gray area indicates the region between the 25th and 75th percentile. 
    Each data point consists of 600 random network realizations. }
	\label{fig:networkdensity}
\end{figure}

Fig.~\ref{fig:pred} displays the short term prediction results for each of the networks, which corresponds to the attractors in Fig.~\ref{fig:lorenz_readouts}. 
From Fig.~\ref{fig:pred}, the predicted signal of the low density network (dotted red) appears to follow closely to the actual Lorenz signal (black), with a forecast time of 4.9 Lyapunov times. This is surpassed with a forecast time of 6.3 Lyapunov times for the medium density network (dashed blue). 
The densest network (dash-dotted green) was unable to learn the Lorenz dynamics, only achieving a forecast time of 0.7 Lyapunov times. 
Note that these results were selected from the best of 1000 random realizations of each network. 

Despite the low density network capable of performing well in short-term autonomous forecasting, it has not sufficiently learned the Lorenz attractor dynamics to predict its long-term behavior, as observed in Fig.~\ref{fig:lorenz_readouts}(b) where the predicted attractor diverges from the true Lorenz attractor after some time. 
The high density network has not learned sufficiently to replicate the Lorenz system dynamics in the short or long term.
The medium density network is the only example here capable of performing good short-term forecasting, while also maintaining the long-term behavior of the Lorenz attractor.

Fig.~\ref{fig:networkdensity} plots the average forecast times with respect to network density on 500-node networks, each average obtained from 600 random realizations of the network.
For direct comparison, the network densities of the low, medium and high density networks in Fig.~\ref{fig:lorenz_readouts} and similar results are $1.7$\%, $8$\% and $99$\% respectively.
Note that due to the effectively unpredictable nature of chaos, there may exist a wide range of forecasting results even for the same overall RC setup, resulting in large standard deviation of forecast times; the autonomous forecasting times can also be improved with larger networks and optimization of other relevant parameters \cite{Zhu2023_L2L}, but for the purpose of comparing reservoir dynamics, a 500 node network with the current relatively simple implementation is sufficient.
Consistent with similar previous results on other tasks \cite{zhuInformation2021}, the better performing networks tend to be skewed toward lower densities, with performance noticeably decreasing above a certain network density.

The complex coupled interplay between the network structure and memristive edge dynamics is the unique property of these physical neuromorphic networks that differentiates them from other RC approaches. 
In conventional algorithmic RC, neuron activations $\mathbf{x}(t+\Delta t)$, in their simplest form, are applied as a mathematical activation function $f$ to a vector constructed from a series of matrix multiplications with the reservoir weight matrix $W$, input weight matrix $W_{\rm in}$ and input signal $\mathbf{u}(t)$ according to
\begin{equation}
    \mathbf{x}(t+\Delta t) = f(W\mathbf{x}(t)+W_{\rm in}\mathbf{u}(t)).
\end{equation}
From the model of physical neuromorphic networks based on memristive nanowire networks \cite{kuncicNeuromorphic2020a,hochstetterAvalanches2021,zhuInformation2021,Baccetti2024}, an analogous expression can be derived as,
\begin{equation}
    \mathbf{x}(t+\Delta t) = M(t)W_{\rm in} \mathbf{u}(t), \label{eq:M}
\end{equation}
where $M(t) = M(t; \mathbf{x}(t); \mathbf{u}(t))$ is a matrix that encompasses all conductance weight updates, Kirchhoff's laws and network structure.
Note that $M(t)$ in \eqref{eq:M} is not a network weight matrix and likewise the conductance of the memristive junction edges in the neuromorphic network cannot be directly interpreted as weights, although for comparison they are the closest in functionality. 

In contrast to algorithmic RC, it is clear from \eqref{eq:M} that the main driving force behind the functionality of neuromorphic reservoirs lies in its dynamical properties. 
Instead of a mathematical activation function and a mathematical recurrent neural network, all nonlinearities arise from the intrinsic dynamics of the nano-electronic circuit elements (memristors) and their coupling to the underlying physical substrate (e.g. nanowire network).
In conventional RC, the reservoir network is static and tends to be fully connected (although recent approaches consider sparsity -- e.g. \cite{Manneschi2023}), which can be advantageous in their mathematical simplicity and usage algorithmically. 
However, physical dynamical systems cannot be mapped onto the same RC scheme because the reservoir is itself dynamic, even though the underlying network structure (the physical nanowire network in this case) is static.

\section{Conclusion}
\label{sec:conclude}
The results of this study demonstrate how the complex nonlinear dynamics inherent in physical reservoirs can be harnessed for physical reservoir computing and how the approach differs from conventional reservoir computing with mathematical reservoirs.
For neuromorphic reservoirs comprised of self--organized nano--electronic networks, in particular, these results reveal how physics constraints couple the underlying physical network structure with nonlinear node--edge (neuro--synaptic) dynamics, 
which places a physical limit on the collective reservoir dynamics. 
This is in contrast to conventional reservoir computing, where nonlinear dynamics are introduced only on the nodes, while edges are kept static, typically on a random or fully--connected network. 

The results from studying physical neuromorphic networks reveal that sparser networks exhibit higher local activity, but less global activity.
In contrast, denser networks display minimal dynamics with low diversity in readouts due to the indiscriminate distribution of voltages, which are used as readouts.
Networks of intermediate density exhibit the most complex readouts and dynamical network activity, optimally forecasting both short-term and long-term behaviors of the Lorenz63 chaotic system. 
Sparse networks can only forecast the short-term evolution, while a dense network is unable to learn the Lorenz system, revealing the importance of appropriate network sparsity in producing dynamical features that embed the salient low-dimensional temporal patterns needed to learn the chaotic attractor.

The neuromorphic network's distinctive functionality arises from its operation as a complex network of dynamical systems. 
Unlike conventional reservoir computing with its usage of activation functions, the neuromorphic network's nonlinear behavior and overall functionality stem from its \textit{in situ} physics--constrained dynamics.

\section*{Acknowledgment}
The authors acknowledge use of the Artemis high performance computing resource at the Sydney Informatics Hub, a Core Research Facility of the University of Sydney. 
Y.X. is supported by an Australian Government Research Training Program (RTP) Scholarship.

\bibliographystyle{ieeetr} % abbreviate and list in citing order
\bibliography{references}

\end{document}